\documentclass[twocolumn,showpacs,aps, prb, floatfix, amsmath]{revtex4}
\usepackage{dcolumn}
\usepackage{bm}
\usepackage{graphicx}
\usepackage{bm}

\begin{document}
\title{Concentration dependence in kinetic arrest of first order magnetic transition in Ta doped $HfFe_2$}
\author{R. Rawat$^a$, P. Chaddah$^a$, Pallab Bag$^a$, P. D. Babu$^b$ and V. Siruguri$^b$}
\affiliation{$^a$UGC-DAE Consortium for Scientific Research, University Campus, Khandwa Road, Indore-452001, India.}
\affiliation{$^b$UGC-DAE Consortium for Scientific Research Mumbai Centre, R-5 Shed, BARC, Mumbai-400085, India.}

\begin{abstract}

Magnetic behavior of the pseudo-binary alloy $Hf_{1-x}Ta_xFe_2$ has been studied, for which the zero field ferromagnetic (FM) to antiferromagnetic (AFM) transition temperature is tuned near to $T=0$ K. These alloys show anomalous thermomagnetic irreversibility at low temperature due to kinetic arrest of the first order AFM-FM transition. All the three studied compositions show re-entrant transition in zero field cooled warming curve and anomalous non-monotonic variation of upper critical field in isothermal magnetization.  The region in H-T space, where these features of kinetic arrest manifest themselves, increases with increasing Ta concentration.    

\end{abstract}



\maketitle
\section{Introduction}
		
	The compound $HfFe_2$ has been reported to show polymorphism with one of the phases crystallizing in $MgZn_2$ type hexagonal structure (space group $P6_3/mmc$) \cite{Nishi1982, Nishi1983}. This hexagonal phase is reported to be ferromagnetic (FM) at room temperature \cite{Cavor2008}. Magnetism in this system arises mainly due to Fe which has two inequivalent sites: 6h and 2a. With Ta substitution for Hf, this phase becomes antiferromagnetic (AFM)  at room temperature and shows AFM to FM transition at lower temperatures \cite{Nishi1983}.  With increasing Ta concentration, the zero field first order AFM-FM transition temperature decreases and is found to be absent for more than 25\% Ta substitution \cite{Nishi1983, Kido1986}. Beyond 70\% Ta substitution, this system is reported to be a Pauli paramagnet. The first order AFM-FM transition for less than 25\% Ta substitution is accompanied with large change in lattice volume. Interestingly, the low temperature FM state has higher volume compared to high temperature AFM state \cite{Kido1986, Morellon1986}. Morellon et al. \cite{Morellon1986} have observed a volume magnetostriction of 0.6\% at low temperature in $Hf_{1-x}Ta_xFe_{2-y}$ with x= 0.15, 0.17 and y=0.2. The Mossbauer studies of Nishihara et al. \cite{Nishi1983}on $Hf_{1-x}Ta_xFe2$ (x= 0 - 0.7) showed discontinuous change in hyperfine field of Fe across first order AFM-FM transition. The hyperfine fields on 6h and 2a Fe sites are of comparable value in FM state, whereas they are reduced drastically in AFM state. Their estimates showed that Fe moments at 6h and 2a sites are 1 $\mu_B$ each in FM state, and 0.7 and 0, respectively, in AFM state. Neutron diffraction study of Duijn et al. \cite{Duijn} confirmed the magnetic moment of 1 $\mu_B$ in FM state but with moment direction lying within the basal plane. The magnetic structure in the AFM state is considered to be similar to isostructural $TiFe_2$ which consists of sublattice of ferromagnetic Fe (6h) layers aligned antiparallel to each other along c-axis with no moment on 2a site \cite{Nishi1983, Duijn}. Delyagin et al. \cite{Delyagin} have proposed canted ferromagnetic structure to interpret their Mossbauer studies on Ti doped $HfFe_2$. However, studies related to the determination of magnetic structure in these system are far from complete. The first order AFM-FM transition in this system appeared to be similar to that observed in FeRh system, e.g., large volume change across first order transition and vanishing of Rh moment in the AFM state \cite{Shirane, Moruzzi}. However, in case of $HfFe_2$, higher volume FM state is the low temperature state whereas for FeRh, lower volume AFM state is the ground state. In contrast to extensively investigated FeRh system, there have been few studies on first order transtion in Ta doped $HfFe_2$ system. Nishihara et al. \cite{Nishi1983} applied the model of itinerant electron magnetism and found a qualitative agreement with it. Wada et al. \cite{Wada} also used this framework to explain higher linear contribution to specific heat at low temperature in antiferromagnetic system. This analysis indicates comparable strength of FM and AFM interactions and measurement under pressure and magnetic field shows that transition temperature can be varied over a wide temperature range \cite{Kido1986, Morellon1986}. 
	
	Such first order magnetic transitions where transition temperature can be shifted to lower temperature have been of interest due to possibility of observing glass like arrest of kinetics, see e.g. \cite{RoyActa, Chaddah2010}. Recently, Chaddah and Banerjee \cite{Chaddah2012} have proposed that magnetic glasses form in these systems when magnetic latent heat is weakly coupled to thermal conduction process. In the present system, first order transformation is accompanied with significant latent heat \cite{Wada} and transition temperature can be varied over a wide temperature range \cite{Nishi1983}. Therefore, such glass like magnetic state can be expected in the present system. 	Here, we present our results on Ta doped $HfFe_2$ for which first order transition temperature is tuned to near $T=0$ Kelvin. Earlier studies have shown that such a composition lies in the range of x=0.20 and 0.25 in $Hf_{1-x}Ta_xFe_2$ \cite{Nishi1983, Kido1986}. Therefore, we prepared samples with Ta concentration (x) varying from 0.15 to 0.30. We find that lattice parameters $a$ and $c$ decrease linearly by about 0.5\% over this range, in accordance with Vegard's Law. Detailed studies for x=0.225, 0.230, and 0.235 compositions show thermomagnetic irreversibility. By following novel paths in H-T space, we show that AFM state is the non-equilibrium state even though the first order transition to FM state is absent in these systems for zero field cooling. The lattice structure for all the studied compositions does not change and there is a slight contraction of less than 0.03\%, allowing us to speculate in the Discussion that the observed changes can be understood as arising from change in density.

\maketitle\section{Experimental details}
Polycrystalline samples of $Hf_{1-x}Ta_xFe_2$ (x=0.225, 0.230 and 0.235) were prepared by arc melting the constituent elements under inert 
Argon gas atmosphere. The purity of Fe and Ta was better than 99.99 \% and Hf purity was 99.9\% (with $\approx$ 2\% of Zr). As prepared 
samples were cut using slow speed diamond saw for x-ray diffraction and magnetic measurements. Powder x-ray diffraction patterns of these 
samples were analyzed using FullProf refinement. Obtained x-ray powder diffraction pattern (symbol) along with fitted pattern (line) are 
shown in figure 1 for these compositions. The refinement shows that these alloys crystallize in $P6_3/mmc$ hexagonal structure and are single phase. Magnetization measurements were performed using a 7 Tesla SQUID-VSM and a 9 Tesla PPMS-VSM, both from Quantum Design, USA.

\begin{figure}[hbt]
	\begin{center}
		\includegraphics[width=7.5 cm]{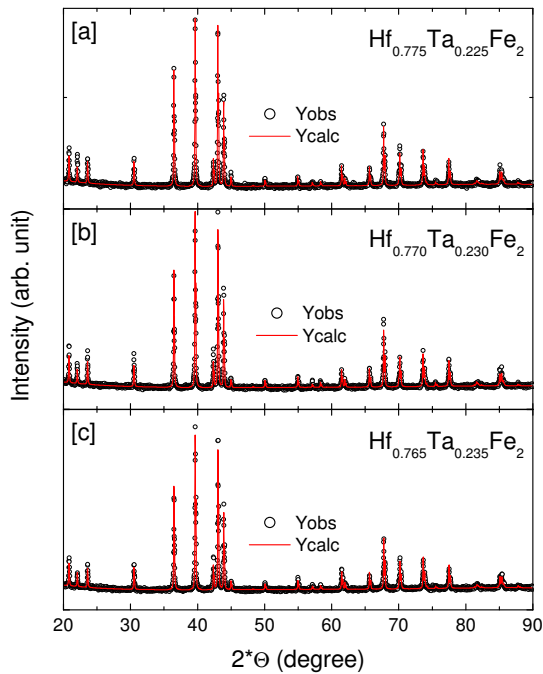}
	\end{center}
	\caption{(color on line) Powder x-ray diffraction pattern (Black circle) along with fitted curve (line) for $Hf_{1-x}Ta_xFe_2$. }
	\label{Figure1}
\end{figure}

\begin{figure}[hbt]
	\begin{center}
		\includegraphics[width=8.0 cm]{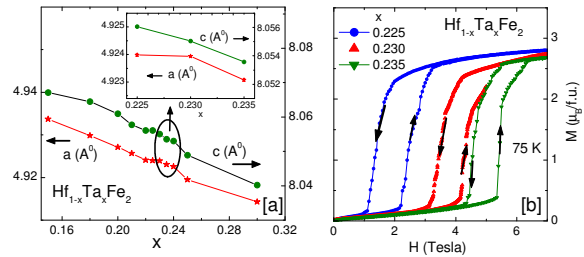}
	\end{center}
	\caption{(color on line) [a] Unit cell parameters $a$ and $c$ for $Hf_{1-x}Ta_xFe_2$. Inset highlights the lattice parameter variation for compositions presented here. [b] Isothermal magnetization at T= 75 K.}
	\label{Figure2v1}
\end{figure}

\begin{figure*}[t]
	\begin{center}
		\includegraphics[width=16 cm]{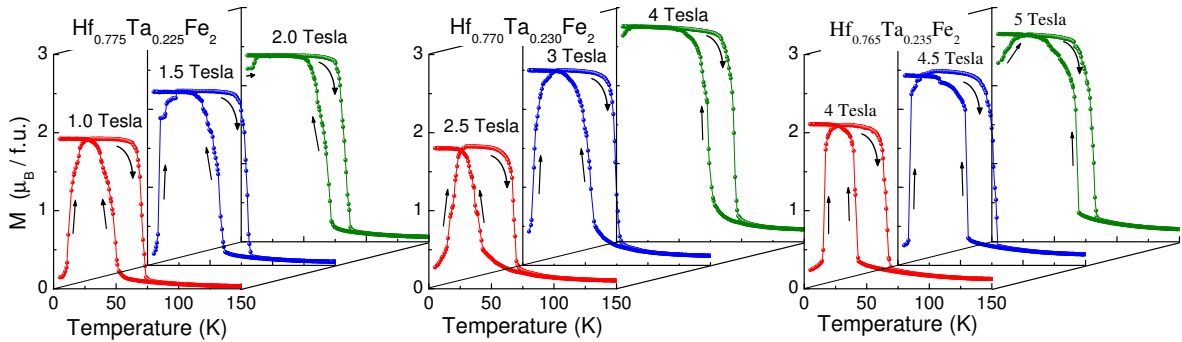}
	\end{center}
	\caption{(color on line) Temperature dependence of magnetization measured under labeled magnetic fields for $Hf_{1-x}Ta_xFe_2$. For each field value, there are two curves ZFCW and FCC. These curves highlight multivalued M well below the transition region of corresponding FCC curve, and reentrant transition in ZFCW.}
	\label{Figure3}
\end{figure*}

\maketitle\section{Results}

Figure 2 [a] shows the unit cell parameters $a$ and $c$ obtained from FullProf refinement of powder XRD patterns. As expected, obtained lattice parameters decrease monotonically with increasing Ta content, following Vegard's law. Inset highlights the variation of lattice parameters for compositions which will be investigated in detail in this manuscript. Figure 2(b) shows the isothermal MH curve at 75 K, which shows that all the three samples are in AFM state at zero field and become FM with the application of magnetic field. Higher the Ta content higher is the magnetic field required for AFM-FM transition. This monotonic increase of critical field with increasing Ta content is consistent with the existing understanding that AFM state becomes more favorable with x.
 
Figure 3 shows temperature dependence of magnetization (M) for x= 0.225, 0.230 and 0.235 in the presence of labeled magnetic field. For each of these field values, magnetic field is applied isothermally at 5 K after zero field cooling and measurement is performed during warming up to 150 K (ZFCW) followed by cooling (FCC). The warming and cooling curves show a hysteretic first order AFM-FM transition for all the three samples. It is to be noted here that for higher x, applied magnetic field is higher which is consistent with MH curves shown in figure 2. The transition temperature increases with increasing magnetic field, as expected for a first order transition from low temperature FM to high temperature AFM state. The ZFCW curve shows a reentrant transition in that the low-T state has a low M (attributed to arrested AFM state) that increases with rising T (attributed to de-arrest of the AFM state to the equilibrium FM state) and again decreases at a much higher T as the FM state undergoes a first order transition to the AFM state. This entire behavior is very similar to the behavior that has been observed in $Gd_5Ge_4$ \cite{Roy2006}, La-Pr-Ca-Mn-O \cite{Wu2005} and $Pr_{0.5}Ca_{0.5}Mn_{0.975}Al_{0.025}O_3$ \cite{Banerjee2006}. Such reentrant transitions have been a hallmark of glass like magnetic states which have been shown to arise due to arrested kinetics of first order transition at low temperature \cite{Kushwaha2009, Roy2006, Banerjee2006, Manekar2001, Chatto2005, Kumar2006, Kushwaha2008, Banerjee2009}. Banerjee et al. \cite{Banerjee2009} have shown that, by following CHUF (cooling and heating in unequal magnetic field) protocol, one can determine which of the two states is the equilibrium state. Though not shown here, we also measured the M-T curve during warming after cooling in a higher field and such curves showed only one transition. Therefore, as reentrant transition is observed during warming only when cooling field is lower, it shows that low field state (here, AFM) is the non-equilibrium arrested state and the equilibrium state obtained on de-arrest is the FM state. The difference between ZFCW and FCW curves decreases for higher field values due to magnetic field induced AFM to FM transition at 5 K, resulting in higher FM phase fraction at 5 K for ZFCW curve. The regions whose kinetic arrest lines are encountered at lower field are transformed,  whereas remaining regions remain arrested. Such regions will show devitrification on warming. Since at higher field values, devitrification curve merges to FCW curve at lower temperature and vice versa, it shows that regions which remains arrested at higher field show dearrest at lower temperature. This conforms to anticorrelated kinetic arrest and supercooling bands \cite{Kumar2006}.

\begin{figure*}[t]
	\begin{center}
		\includegraphics[width=15.5 cm]{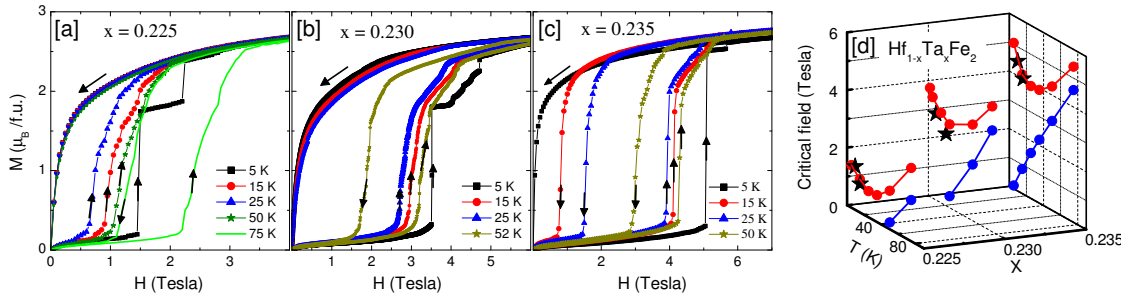}
	\end{center}
	\caption{(color on line) Isothermal magnetization measured after zero field cooling for $Hf_{1-x}Ta_xFe_2$. Arrows indicate field sweep direction. These curves show anomalous behavior for forward MH curve (0 to 8 Tesla) which shifts to higher field side above and below 25 K.}
	\label{Figure4}
\end{figure*}

For a given composition, it can be seen that magnetization below the hysteresis region of first order transition (in FCC) is higher for higher applied field e.g. for x= 0.230 and 5 K, it is 1.8, 2.5 and 2.8 $\mu_B/f.u.$ for 2.5, 3 and 4 Tesla applied magnetic fields, respectively. This could arise either due to increasing FM phase fraction or due to susceptibility of ferromagnetic phase. To verify this, we measured isothermal magnetization at various temperatures. For these measurements, sample is cooled to measurement temperature and MH is measured during increasing and decreasing field cycle, i.e. 0-7-0 Tesla. These curves are shown in figure 4 and to highlight critical field (field required for AFM to FM and FM to AFM transition) variation with temperature, data is shown for some representative temperatures only. The critical field (which we have taken as the magnetic field at which M reaches half of the value at 7 Tesla for respective temperature) required for AFM to FM transition $H_{up}$ is higher for higher Ta content for a given temperature. For example, at 5 K, it is observed to be 1.48, 3.48 and 5.06 Tesla for x=0.225, 0.230 and 0.235, respectively. Extrapolation of these critical field values is in good agreement with earlier study \cite{Kido1986}. The return curve (7 to 0 Tesla) varies smoothly without any clear signature of FM to AFM transition  and becomes zero at zero field value. This decrease in M at low field could be due to soft ferromagnetic nature of the FM phase or due to back transformation of FM phase to AFM phase. To check these possibilities, we measured MH in subsequent third cycle (0 to 7 Tesla) to obtain envelope hysteresis curve at 5 K. In all the three cases, virgin curve lies outside the envelop curve (not presented here for the sake of clarity) showing that FM phase is retained down to zero field. This also rules out the magnetocrystalline anisotropy as the probable cause of increase in M for ZFCW curve shown in figure 1. At higher temperature, the return curves (7 to 0 Tesla)  move to higher H, which is expected for a first order transition from low temperature ferromagnetic state to high temperature antiferromagnetic state. For such cases, transition temperature (critical field) increases with increasing magnetic field (temperature).  In contrast, forward curves (0 to 7 Tesla) show anomalous behavior, i.e, they shift to lower H with increasing temperature and above 25-35 K, they start to shift to higher H. Such anomalous shift in MH curve with temperature has been observed in the return curve of MH in $Nd_{0.5}Sr_{0.5}MnO_3$ \cite{Rawat2007} and RH curve of Pd doped FeRh \cite{Kushwaha2009} and has been attributed to kinetic arrest of first order transition at low temperature. This anomalous feature becomes more obvious when we draw a phase diagram using these isothermal MH measurements.  

The phase diagram based on these isothermal M-H measurement is shown in figure 4[d]. It shows that with increasing Ta content, FM region shifts to lower temperature and higher field. The lower critical field ($H_{dn}$) increases monotonically with increasing temperature whereas the upper critical field ($H_{up}$) shows non-monotonic variation for all the compositions with minima around 25-35 K. Such non-monotonic variation in critical field has been shown to be a consequence of interplay between kinetic arrest line and supercooling line \cite{Rawat2007}. At low temperature, kinetics dominate whereas at high temperature, thermodynamics takes over. In case of $Gd_5Ge_4$ \cite{Roy2006} and $La-Pr-Ca-Mn-O$ \cite{Wu2005} also, a similar non-monotonic variation of $H_{up}$ has been shown to give rise to glass like AFM state at low temperature and field. The low temperature (below $\approx$ 25 K) side of $H_{up}$ in H-T space then represents kinetic arrest $(H_K, T_K)$ band, below which AFM to FM transition is hindered on experimental time scales. In contrast, the high temperature side represents the supercooling band. Since these two bands have opposite slope, there results a minima in $H_{up}$ at intermediate temperatures.

These phase diagrams also explain the re-entrant transition observed in ZFCW curve, as shown in figure 1. For lower field value, kinetic arrest line is crossed before crossing the supercooling line and therefore, system remains in AFM state. When the magnetic field is applied after zero field cooling, it partly or completely transforms to AFM state. The part transformation occurs due to quench disorder broadening of first order transition line where different regions on the length scale of correlation length can have different transition temperatures.  On warming in the presence of field when $(H_K,T_K)$ line is crossed for remaining AFM phase, it transforms to FM state. Therefore, AFM to FM transformation for ZFCW curve occurs across kinetic arrest line. From these devitrification curves, kinetic arrest temperature is estimated as the temperature at which magnetization value reaches half of the saturation magnetization. These points, shown by black stars in the phase diagram, coincide with the rising part of $H_{up}$ curve at low temperature.

The phase diagram shown in figure 4(d) also shows that $T_K(H)$ rises monotonically as $x$ varies from 0.225 to 0.235. The formation of a magnetic glass is thus found to become more favorable as $x$ increases. As noted earlier, the lattice parameters decrease monotonically as $x$ increases. This would favor a direct exchange magnetic coupling, as the wave function overlap would rise exponentially, over a conduction electron mediated coupling (that would not vary much with interatomic spacing). Therefore, our results could be relevant to the proposal of Chaddah and Banerjee \cite{Chaddah2012} that a magnetic glass is formed when the magnetic latent heat is weakly coupled (c.f. the sample specific heat) to the thermal conduction process. 

\maketitle\section{Conclusions}

To conclude, we prepared and studied the magnetic behavior of the pseudo-binary alloys $Hf_{1-x}Ta_xFe_2$ for which transition temperature is tuned to $T=0$ K. These alloys showed anomalous thermomagnetic irreversibility at low temperature due to kinetic arrest of the first order AFM-FM transition. All the three studied compositions showed re-entrant transition in ZFCW, anomalous variation of forward curve in isothermal magnetization and non-monotonic variation of upper critical field. With increasing Ta concentration, the AFM state is stabilized and higher magnetic field is required to induce the FM state isothermally. The region of co-existing AFM and FM phases increases in H-T space with increasing Ta concentration. These measurements clearly show that for lower $x$ values, the ground state is FM even though the zero field cooled state is AFM. Due to the interplay of kinetic arrest and supercooling, tunable fractions of AFM and FM phases can be obtained for the same temperature and magnetic field values. Such contrasting magnetic states for same thermodynamic parameters will be helpful to understand the magnetic structure of these systems. This understanding can provide further insight on the first order AFM-FM transition in FeRh system where transition is found to be sensitive to particle size \cite{Navarro1996}, strain \cite{Fan2010} etc. Whereas in case of Ta doped $HfFe_2$, the first order transition is observed even in powder sample as is evident from the extensive Mossbauer studies in these systems.

\maketitle\section{Acknowledgments}

We acknowledge M. Gupta for XRD measurements and Komal Bapna, R. J. Choudhary for magnetic measurements with SQUID-VSM.


\begin{thebibliography}{0}
\expandafter\ifx\csname natexlab\endcsname\relax\def\natexlab#1{#1}\fi
\expandafter\ifx\csname bibnamefont\endcsname\relax
  \def\bibnamefont#1{#1}\fi
\expandafter\ifx\csname bibfnamefont\endcsname\relax
  \def\bibfnamefont#1{#1}\fi
\expandafter\ifx\csname citenamefont\endcsname\relax
  \def\citenamefont#1{#1}\fi
\expandafter\ifx\csname url\endcsname\relax
  \def\url#1{\texttt{#1}}\fi
\expandafter\ifx\csname urlprefix\endcsname\relax\def\urlprefix{URL }\fi
\providecommand{\bibinfo}[2]{#2}
\providecommand{\eprint}[2][]{\url{#2}}

\end{thebibliography}


\begin{thebibliography}{}

\bibitem[1]{Nishi1982} Y. Nishihara and Y. Yamaguchi, J. Phys. Soc. Jpn. \textbf{51} (1982) 1333.
\bibitem[2]{Nishi1983}  Y. Nishihara and Y. Yamaguchi, J. Phys. Soc. Jpn. \textbf{52} (1983) 3630.
\bibitem[3]{Cavor2008} J. Belosevic-Cavor, V. Koteski, N. Novakovic, G. Concas, F. Congiu and G. Spano, Eur. Phys. J. B \textbf{50} (2006) 425.
\bibitem[4]{Kido1986} G. Kido, Y. Tadakuma, Y. Nakagawa, Y. Nishihara and Y. Yamaguchi, J. Mag. Mag. Mater. \textbf{54-57} (1986) 885.
\bibitem[5]{Morellon1986} L. Morellon, P. A. Algarabel and M. R. Ibarra, Z. Arnold and J. Kamarad, J. Appl. Phys. \textbf{80}, 6911 (1996).
\bibitem[6]{Duijn} H. G. M. Duijn, E. Bruck, A. A. Menovsky, K. H. J. Buschow, F. R. de Boer, R. Coehoorn, M. Winkelmann and K. Siemensmeyer, J. Appl. Phys. \textbf{81} (1997) 4218.
\bibitem[6]{Delyagin} N.N. Delyagin, A.L. Erzinkyan, V.P. Parfenova, I.N. Rozantsev, G.K. Ryasny, J. Mag. Mag. Mater. \textbf{320} (2008) 1853.
\bibitem[7]{Shirane} G. Shirane, C. W. Chen, P. A. Flinn, and R. Nathans, J. Appl. Phys. \textbf{34} (1963) 1044; G. Shirane, R. Nathans, and C. W. Chen, Phys. Rev. \textbf{134} (1964) A1547.
\bibitem[8]{Moruzzi} V. L. Moruzzi and P. M. Marcus, Phys. Rev. B \textbf{46} (1992) 2864.
\bibitem[9]{Wada} H. Wada, N. Shimamura, and M. Shiga, Phys. Rev. B \textbf{48} (1993) 10221.
\bibitem[10]{RoyActa} S.B. Roy, P. Chaddah, V.K. Pecharsky, K.A. Gschneidner Jr., Acta Materialia \textbf{56} (2008) 5895
\bibitem[11]{Chaddah2010} P Chaddah and A Banerjee arXiv:1107.0125v1.
\bibitem[12]{Chaddah2012} P Chaddah and A Banerjee arXiv:1004.3116v3.
\bibitem[13] {Roy2006} S. B. Roy, M. K. Chattopadhyay, P. Chaddah, J. D. Moore, G. K. Perkins, L. F. Cohen, K. A. Gschneidner, Jr., and
V. K. Pecharsky,  Phys. Rev. B \textbf{74} (2006) 012403.
\bibitem[14] {Wu2005} W. Wu, C. Israel, N. Hur, P. Soonyong, S. W. Cheong and A. de Lozanne, Nat. Matter. 5 (2006) 881.
\bibitem[15] {Banerjee2006} A. Banerjee, K. Mukherjee, Kranti Kumar, and P. Chaddah, Phys. Rev. B \textbf{74} (2006) 224445.
\bibitem[16] {Kushwaha2009} Pallavi Kushwaha, Archana Lakhani, R. Rawat and P. Chaddah, Phys. Rev. B \textbf{80} (2009) 174413.
\bibitem[17] {Manekar2001} M. A. Manekar el al., Phys. Rev. B \textbf{64} (2001) 104416.
\bibitem[18] {Chatto2005} M. K. Chattopadhyay, S. B. Roy, and P. Chaddah, Phys. Rev. B \textbf{72} (2005) 180401R.
\bibitem[19]{Kumar2006} Kranti Kumar, A.K. Pramanik, A. Banerjee, P. Chaddah, S.B. Roy, S. Park, C.L. Zhang and S. W. Cheong, Phys. Rev. B \textbf{73} (2006) 184435. 
\bibitem[20]{Kushwaha2008} Pallavi Kushwaha, R. Rawat, and P. Chaddah, J. Phys.: Condens Mater \textbf{20} (2008) 022204
\bibitem[21]{Banerjee2009} A. Banerjee, A.K. Pramanik, Kranti Kumar and P. Chaddah, J. Phys.: Condens. Matter \textbf{18} (2006) L605; A. Banerjee, Kranti Kumar and P. Chaddah, J. Phys.: Condens. Matter \textbf{21} (2009) 026002. 
\bibitem[22]{Rawat2007} R. Rawat, K. Mukherjee, Kranti Kumar, Alok Banerjee and P Chaddah, J. Phys.: Condens. Matter \textbf{19} (2007) 256211.
\bibitem[23]{Navarro1996} E. Navarro, M. Multigner, A. R. Yavari and A. Hernando, Europhys. Lett. \textbf{35} (1996) 307-311.
\bibitem[24]{Fan2010} R. Fan et al. Phys. Rev. B 82 (2010) 184418; C. Bordel et al., Phys. Rev. Lett \textbf{109} (2012) 117201.

\end{thebibliography}
\end{document}